\documentclass[twocolumn,showpacs,prd,aps,tightenlines,floatfix,final]{revtex4}

\pdfoutput=1

\usepackage[latin1]{inputenc}

\usepackage{graphics,graphicx}

\usepackage{color}
\usepackage{amsmath,amssymb}

\usepackage[percent]{overpic}
\usepackage{rotating}

\usepackage{array}
\newcolumntype{x}[1]{%
>{\centering\hspace{0pt}}p{#1}}%

\usepackage{warpcol}
\newcolumntype{L}[1]{>{\pcolbegin{r}{#1}}l<{\pcolend}}
\newcolumntype{R}[1]{>{\pcolbegin{r}{#1}}r<{\pcolend}}

\def \dd{\mathrm{d}}
\newcommand{\parder}[2]{\frac{\partial #1}{\partial #2}}
\newcommand{\bs}[1]{\boldsymbol{#1}}

\begin{document}

\title{Symplectic Integration of Post-Newtonian Equations of Motion with Spin}

\author{Christian~Lubich$^1$, Benny~Walther$^2$, Bernd~Br\"{u}gmann$^2$}

\affiliation{$^1$Mathematics Institute, University of T\"{u}bingen, 72076 T\"{u}bingen, Germany\\
$^2$Theoretical Physics Institute, University of Jena, 07743 Jena, Germany}

\date{March 26, 2010}

\begin{abstract}
We present a non-canonically symplectic integration scheme tailored to numerically computing the post-Newtonian motion of a spinning black-hole binary. Using a splitting approach we combine the flows of orbital and spin contributions. In the context of the splitting, it is possible to integrate the individual terms of the spin-orbit and spin-spin Hamiltonians analytically, exploiting the special structure of the underlying equations of motion. The outcome is a symplectic, time-reversible integrator, which can be raised to  arbitrary order by composition. A fourth-order version is shown to give excellent behavior concerning error growth and conservation of energy and angular momentum in long-term simulations. Favorable properties of the integrator are retained in the presence of weak dissipative forces due to radiation damping in the full post-Newtonian equations.
\end{abstract}


\maketitle


\section{Introduction}
\label{sec:Introduction}
\vskip-0.3cm

In general relativity there is a rich phenomenology associated with
two orbiting compact objects and the gravitational waves that are
emitted in the process. These phenomena can be studied to very high
accuracy in the post-Newtonian approximation of general relativity,
see e.g.~\cite{Bla06}. The post-Newtonian equations of motion are
well-known up to a certain order, and numerical integration of the
resulting ordinary differential equations can be performed to obtain
the orbits without immediate difficulty, at least as long as certain
evolution times and accuracy requirements are not exceeded.

In this paper we address the question of long-term integration of the
post-Newtonian equations of motion, allowing unequal masses and
spins. A study of a large number of
orbits is of interest for various reasons, of which we want to mention
only one. In general relativity, there is the possibility of chaotic
orbits, which can place severe demands on the quality of the numerical
integrator.

The occurrence of chaotic trajectories of orbiting spinning binaries
has been discussed, for example,
in~\cite{Cor01,Lev03,KiuMaeKei04,GopKoe05,WuXie08}, implying
unpredictable irregularities in the waveforms. However, there still
seems to be a controversy among different authors under what
conditions the post-Newtonian equations lead to chaotic motion.
Various possible indicators for chaos like Lyapunov exponents,
Poincar\'{e} sections, and basin boundary plots, are employed in the
investigations found in the literature, and interesting regions of the
configuration space, including mass ratio, eccentricity, spin
orientations and lengths, are densely covered by thousands of
numerical simulations. Typical studies of chaos require
numerical long-term evolutions of the system, depending on the
timescale at which chaotic behavior becomes apparent.

The structure of the post-Newtonian equations of motion permit a
separate investigation of bodies orbiting in a conservative fashion
(not incorporating radiation damping) and of inspiraling masses losing
energy by emission of gravitational waves.  The conservative case
facilitates the detection of chaotic tendencies as simulations can in
principle be carried out arbitrarily long, whereas in the dissipative
scenario the objects would finally merge.
Both kinds of simulations, conservative and non-conservative, would
benefit from an efficient and well-behaved long-term integrator.

Computational challenges from a wide variety of areas of research like
celestial mechanics, molecular dynamics, quantum mechanics, as well as
abstract numerical analysis motivated the development of
structure-preserving algorithms for differential equations. Geometric
integrators such as symplectic or symmetric methods are known, from
numerical experience and theory, to yield a substantially improved
long-term behavior compared to standard integrators such as explicit
Runge-Kutta methods. The improved behavior concerns the exact or
approximate preservation of conserved quantities (like energy and
angular momentum) without drift, slower error growth, and also a more
faithful representation of Poincar\'e sections as used in
investigations of chaotic behavior. We refer to the monographs
\cite{HaiLubWan06,LeiRei04} and numerous references therein.

In the following we construct a structure-preserving, efficiently
implementable integrator for the equations of motion of binary
spinning black holes in the post-Newtonian approximation.  The integrator
is based on a splitting into conservative motion and dissipative
perturbation due to radiation-reaction forces, into orbital and spin
evolution, and on further splittings between the Newtonian and
post-Newtonian Hamiltonian and between different terms in the spin
Hamiltonian. 

In the conservative case, the resulting algorithm is a
Poisson integrator (or non-canonically symplectic integrator in
another terminology) and time-reversible.  The term 'non-canonical'
refers to the spin algebra, which yields a non-canonical Poisson bracket for which no standard structure-preserving integrators exist. 
The Poisson integrator 
preserves the spin lengths exactly. Our numerical experiments show no
drift in energy and total angular momentum.

The paper is organized as follows. In Section II we present the
post-Newtonian Hamiltonian equations of motion in ADMTT gauge
(Arnowitt-Deser-Misner transverse-traceless). The construction of the
symplectic splitting integrator, which exploits the special structure
of that Hamiltonian, is described in Section III. For the conservative
equations the performance of the new integrator is contrasted with the
behavior of the classical Runge-Kutta method in Section IV. Finally,
in Section V we incorporate the dissipative terms into the integrator.


\section{Post-Newtonian equations of motion for spinning binary systems}
\label{sec:pneom}

This section  presents the equations of motion
of a black-hole binary system consisting of objects with masses $m_a$,
positions $\bs{X}_a$, momenta $\bs{P}_a$ and spins $\bs{S}_a$
($a=1,2$). For our purposes it is sufficient to restrict
considerations to the center-of-mass dynamics, where
$\bs{P}\equiv\bs{P}_1=-\bs{P}_2$. 

The Hamiltonian formalism turns out to be very useful when working
with the canonically conjugate position and momentum variables and in
distinguishing conservative and radiative effects. The system's
equations of motion take the form
\begin{align}
 \frac{\dd\bs{X}}{\dd t}&=\left\{\bs{X},H \right\}=\parder{H}{\bs{P}},\label{eq:nonconservx}\\
 \frac{\dd\bs{P}}{\dd t}&=\left\{\bs{P},H \right\}+\bs{F}=-\parder{H}{\bs{X}}+\bs{F},\label{eq:nonconservp}\\
 \frac{\dd\bs{S}_a}{\dd t}&=\left\{\bs{S}_a,H \right\}=\parder{H}{\bs{S}_a}\times\bs{S}_a,\label{eq:spineom}
\end{align}
where $\bs{F}$ is the non-conservative force and $\times$
denotes the usual vector cross product. The conservative part consists
of an orbital and a spin contribution with the composed Hamiltonian
\begin{align}
 H(\bs{X},\bs{P}, \bs{S}_1, \bs{S}_2)=&H_\text{Orb}(\bs{X},\bs{P})+H_\text{Spin}(\bs{X},\bs{P},\bs{S}_1,\bs{S}_2).\label{eq:sumhamiltonian}
\end{align}
Depending on the choice of gauge, different formulations are possible. We use the 3PN accurate orbital Hamiltonian in ADMTT gauge derived by Damour, Jaranowski, Sch\"{a}fer~\cite{DamJarSch99,DamJarSch00b,DamJarSch00c,DamJarSch01}. It forms an expansion in the parameter $1/c^2$ and goes beyond the classical Newtonian Hamiltonian $H_\text{N}$ by adding post-Newtonian, relativistic corrections,
\begin{align}
 H_\text{Orb}(\bs{X},\bs{P})=&\,Mc^2+H_{\text{N}}(\bs{X},\bs{P})
+\frac{1}{c^2}H_{1\text{PN}}(\bs{X},\bs{P})\nonumber\\
&+\frac{1}{c^4}H_{2\text{PN}}(\bs{X},\bs{P})
+\frac{1}{c^6}H_{3\text{PN}}(\bs{X},\bs{P}),\label{eq:pnadmhamiltonian}
\end{align}
where $M=m_1+m_2$ denotes the total rest mass. Although the
convergence of this series can be slow, the magnitude of the energy
and force contributions associated with the individual terms
nevertheless decreases with order. 
However, the complexity of the single terms, whose explicit form can be found in \cite{DamJarSch99,DamJarSch00b,DamJarSch00c,DamJarSch01}, 
increases considerably with order, which seriously affects the computational costs.

The spin contribution in Eq.~(\ref{eq:sumhamiltonian}) also forms a series expansion. We are going to consider the leading-order terms proportional to $G/c^2$ and up to quadratic in the spins only. These expressions can be grouped in spin-orbit interactions and spin-spin interactions~\cite{DamSch88a,Dam01},
\begin{align}
H_\text{Spin}&=H_\text{SO}+H_{S_1S_1}+H_{S_1S_2}+H_{S_2S_2},\label{eq:hspin}
\end{align}
where
\begin{align}
H_{\text{SO}}&=2\frac{G}{c^2}\frac{\bs{S}_\text{eff}\cdot \bs{L}}{R^3}\label{eq:hso},\\
\bs{S}_{\text{eff}}&=\left(1+\frac34 \frac{m_2}{m_1} \right)\bs{S}_1+\left(1+\frac34 \frac{m_1}{m_2}\right)\bs{S}_2,\\[12pt]
H_{S_1S_2}&=\frac{G}{c^2}\frac{1}{R^3}\Big[3\left(\bs{S}_1\cdot \bs{N}\right)\left(\bs{S}_2\cdot \bs{N}\right)-\left(\bs{S}_1\cdot \bs{S}_2\right)\Big],\label{eq:hs1s2}\\
H_{S_1S_1}&=\frac{1}{2}\frac{G}{c^2}\frac{1}{R^3}\Big[3\left(\bs{S}_1\cdot \bs{N}\right)\left(\bs{S}_1\cdot \bs{N}\right)-\left(\bs{S}_1\cdot \bs{S}_1\right)\Big]\frac{m_2}{m_1},\label{eq:hsisi}\\
H_{S_2S_2}&=1\rightleftharpoons 2.
\end{align}
Here, $\bs{L}=\bs{X}\times\bs{P}$ is the orbital angular momentum and
$\bs{N}$ is the unit vector $\bs{X}/R$, where $R=|\bs{X}|$. We assume the Newton-Wigner spin supplementary condition, which provides the notion of the spin vector and affects the definition of the bodies'
worldlines.

Turning towards the radiation-reaction-force term $\bs{F}$ to be applied in
Eq.~(\ref{eq:nonconservp}), we use the expression derived
by~\cite{BuoCheDam05} in cases where we leave the conservative regime. The expression reads
\begin{align}
\bs{F}=&\frac{1}{\omega|\bs{L}|}\frac{\dd E}{\dd t}\bs{P}+\frac{8}{15}\nu^2\frac{v_{\omega}^8}{\bs{L}^2R}\bigg\lbrace\bigg(61+48\frac{m_2}{m_1}\bigg)\bs{P}\!\cdot\!\bs{S}_1\nonumber\\ &+\bigg(61+48\frac{m_1}{m_2}\bigg)\bs{P}\!\cdot\!\bs{S}_2\bigg\rbrace\bs{L},\label{eq:reactionforce}
\end{align}
where $\nu=\mu/M$ with $\mu=m_1m_2/M$ being the reduced mass. It is
crucial to note that the force expression depends on the orbital
angular frequency $\omega$, which is also hidden in the invariant velocity parameter $v_\omega$,
\begin{equation}
v_\omega=\left(\frac{GM\omega}{c^3}\right)^{1/3}.
\end{equation}
The problem with $\omega$ is that it brings the velocity $\dot{\bs{X}}=\dd\bs{X}/\dd t$ into play,
\begin{equation}
\omega=\frac{\vert\dot{\bs{X}}-\bs{N}(\bs{N}\cdot\dot{\bs{X}})\vert}{R},
\end{equation}
which is inconvenient when dealing with momenta. Note that the link between the momentum $\bs{P}$ and the velocity is given by the complicated post-Newtonian relation $\dot{\bs{X}}=\partial H(\bs{X},\bs{P}, \bs{S}_1, \bs{S}_2)/ \partial \bs{P}$. 
The parameter $\omega$ is also needed to express the energy-loss rate $\dd E/\dd t$ appearing in Eq.~(\ref{eq:reactionforce}). We will use the 3.5PN accurate flux for circularly orbiting masses which has been derived by \cite{BlaFayIye02,BlaFayIye05, BlaDamEsp04} and is given in \cite{BuoCheDam05} as
\begin{align}
\frac{\dd E}{\dd t}=&-\frac{32}{5}\nu^2v_{\omega}^{10}\bigg\lbrace 1+f_2(\nu)v_{\omega}^2+[f_3(\nu)+f_{3\text{SO}}]v_{\omega}^3\nonumber\\
&+[f_4(\nu)+f_{4\text{SS}}]v_{\omega}^4+f_5(\nu)v_{\omega}^5+f_6(\nu)v_{\omega}^6\label{eq:taylorflux}\\
&+f_{l6}\,v_{\omega}^6\,\ln(4v_{\omega})+ f_7(\nu)v_{\omega}^7\bigg\rbrace.\nonumber
\end{align}
The coefficients of this series are constant, except for $f_\text{3SO}$ and $f_\text{4SS}$, which depend on the spins and the direction of $\bs{L}$.

It is worth stressing that the energy flux in this form is strictly valid only for
planar, circular motion without any trace of eccentricity. The spin-orbit interaction, $H_\text{SO}$, is known to produce a precession of the orbital plane. While this slow precession is supposed to give only higher-order deviations in the energy flux, the orbital oscillations provoked by the spin-spin terms would lead to noticeable inaccuracies. A more complicated version of the flux which is valid also for eccentric orbits has been derived by \cite{AruBlaIye08}.

However, for our purposes the dissipation according to Eq.~(\ref{eq:taylorflux}) is considered sufficiently accurate. We do not consider it as the goal of this work to evaluate the range of validity of the equations, such as the restriction of Eq.~(\ref{eq:hspin}) to small spin magnitudes, for example, or the complete breakdown of the equations of motions at small binary separations. Instead we want to take the equations, or parts of them, as given and present numerical methods well-suited for these kinds of equations.


\section{A symplectic splitting integrator for the conservative system}
\label{sec:sympsch}
We first focus on the conservative motion, with the dissipative force $\bs{F}$ set to zero. The system (\ref{eq:nonconservx})--(\ref{eq:spineom}) is then a non-canonical Hamiltonian system (or Poisson system) with the Poisson bracket composed of a canonical bracket and the two spin brackets,
\begin{align}
\{ F, G\} = & \left(\parder{F}{\bs{X}}\cdot\parder{G}{\bs{P}}
-\parder{F}{\bs{P}}\cdot\parder{G}{\bs{X}}\right) 
\nonumber
\\ & + \sum_{a=1}^2 \text{det}\left( \parder{F}{\bs{S}_{a}},\bs{S}_{a},\parder{G}{\bs{S}_{a}}\right).
\label{eq:poisson-bracket}
\end{align}
The exact flow $\varphi_t^H:(\bs{X}(0),\bs{P}(0),\bs{S}_1(0),\bs{S}_2(0)) \mapsto
(\bs{X}(t),\bs{P}(t),\bs{S}_1(t),\bs{S}_2(t))$ is a Poisson map
(or non-canonically symplectic map in another terminology):
For all smooth functions $F=F(\bs{X},\bs{P},\bs{S}_1,\bs{S}_2)$ and $G$, the flow preserves the bracket as
\begin{equation}
 \{ F\circ \varphi_t^H,G\circ \varphi_t^H \} = \{ F,G \} \circ \varphi_t^H. 
\end{equation}
In addition to the total energy $H$, the spin lengths $|\bs{S}_a|$ and the total angular momentum $\bs{J}=\bs{L}+\bs{S}_1+\bs{S}_2$ are conserved quantities.

For the numerical treatment of this system we propose a splitting integrator with the following structure-preserving properties (see, e.g.,  \cite{HaiLubWan06} for terminology): 
\begin{itemize}
\item
The method is a {\it Poisson integrator} (or non-canonically symplectic integrator) for the bracket (\ref{eq:poisson-bracket}):
For all smooth functions $F$ and $G$, the discrete flow $\Phi_h^H$ over a stepsize $h$ preserves the bracket,
\begin{equation}
 \{ F\circ \Phi_h^H,G\circ \Phi_h^H \} = \{ F,G \} \circ \Phi_h^H. 
\end{equation}
The method preserves the spin lengths $|\bs{S}_a|$ (the Casimirs of the bracket). These properties imply that a step of the method equals the exact flow 
(up to terms that are exponentially small in the inverse stepsize)
of a Poisson system with the original bracket (\ref{eq:poisson-bracket}) and a slightly modified Hamiltonian (see \cite[Chap.\,IX]{HaiLubWan06}).
\item 
The method is {\it symmetric} and therefore preserves all reversal symmetries present in the system.
\end{itemize}
The basic method is of second order, but it can be enhanced to higher orders by composition.
 
The method is based on a splitting of the Hamiltonian into its orbital, spin-orbit and spin-spin contributions,
\begin{align}
 H=H_{\text{Orb}} + H_{\text{SO}} + H_{\text{SS}},
\end{align}
with further splittings of $H_{\text{Orb}}$, $H_{\text{SO}}$ and $H_{\text{SS}}$ to be described below. Accordingly, we start from the approximation of the flow $\varphi_h^H$ of the system  (\ref{eq:nonconservx})--(\ref{eq:spineom}) with $\bs{F}=0$ over a time step $h$ by the symmetric splitting
\begin{align}
 \varphi_h^H 
&\approx \varphi_{h/2}^{H_{\text{SS}}}\circ\varphi_{h/2}^{H_{\text{SO}}}
\circ 
\varphi_{h}^{H_{\text{Orb}}}\circ
\varphi_{h/2}^{H_{\text{SO}}}\circ \varphi_{h/2}^{H_{\text{SS}}}.
\end{align}
The individual flows in this formula will now be further approximated in a structure-preserving way.

{\bf Orbital integrator.}
For the approximation of the flow of the orbital Hamiltonian
$H_{\text{Orb}}=H_{\text{N}} + H_{\text{PN}}$ we further split into the Newtonian part, which yields pure Kepler motion, and the computationally expensive Post-Newtonian corrections:
\begin{equation}
 \varphi_{h}^{H_{\text{Orb}}}\approx \varphi_{h/2}^{H_{\text{PN}}}\circ \varphi_{h}^{H_{\text{N}}}\circ \varphi_{h/2}^{H_{\text{PN}}}.
\end{equation}
Besides the possibility to use the exact Kepler flow $\varphi_{h}^{H_{\text{N}}}$, we employ a high-order symplectic approximation to it. The sixth-order method labeled p9s9 obtained by symmetric composition of St\"ormer-Verlet steps (see \cite[Sect.\,V.3.2]{HaiLubWan06}) showed best performance in our case.

The post-Newtonian flow $\varphi_{h/2}^{H_{\text{PN}}}$ is approximated by the symplectic Euler scheme 
\begin{align}
\Phi_{h/2}^{H_{\text{PN}}}\,:\quad
\bs{X}_{n+1/2}=&\bs{X}_{n} + \frac{h}{2}\,\parder{H_{\text{PN}}}{\bs{P}}\left(\bs{X}_{n+1/2},\bs{P}_{n} \right)\nonumber
\\
\bs{P}_{n+1/2}=&\bs{P}_{n} - \frac{h}{2}\,\parder{H_{\text{PN}}}{\bs{X}}\left(\bs{X}_{n+1/2},\bs{P}_{n}\right)\label{eq:phipn}
\end{align}
 and its adjoint (exchange $n\leftrightarrow n+1$ and $h\leftrightarrow -h$),
\begin{align}
\Phi_{h/2}^{{H_{\text{PN}}},\ast}:\
\bs{X}_{n+1}=&\bs{X}_{n+1/2} + \frac{h}{2}\,\parder{H_{\text{PN}}}{\bs{P}}\left(\bs{X}_{n+1/2},\bs{P}_{n+1} \right)\nonumber
\\
\bs{P}_{n+1}=&\bs{P}_{n+1/2} - \frac{h}{2}\,\parder{H_{\text{PN}}}{\bs{X}}\left(\bs{X}_{n+1/2},\bs{P}_{n+1}\right).\label{eq:phipnadjoint}
\end{align}
For the non-separable post-Newtonian Hamiltonian $H_{\text{PN}}$, these steps are implicit in $\bs{X}_{n+1/2}$ for 
$\Phi_{h/2}^{{H_{\text{PN}}}}$ and in $\bs{P}_{n+1}$ for
$\Phi_{h/2}^{{H_{\text{PN}}},\ast}$.
The post-Newtonian corrections are expected to be only minor, and so the initial values of $\bs{X}$ and $\bs{P}$ are excellent starting values for the fixed-point iterations.
Our orbital integrator thus approximates the flow $\varphi_h^{H_\text{Orb}}$ by
\begin{align}
\Phi_h^{H_\text{Orb}}=\Phi_{h/2}^{{H_{\text{PN}}},\ast}\circ \Phi_{h}^{H_\text{N}} \circ \Phi_{h/2}^{{H_{\text{PN}}}}.
\end{align}

{\bf Rotations.} For a Hamiltonian
\begin{equation}
 H^{\text{rot}} = \bs{\Omega}\cdot\bs{S}_a
\end{equation}
with a {\it constant} vector $\bs{\Omega}$, the equations of motion 
\begin{equation}
 \dot{\bs{S}}_a = \bs{\Omega}\times \bs{S}_a
\end{equation}
are readily solved by a rotation given by Rodrigues' formula,
\begin{align}
 \bs{S}_a(t)&={\cal{R}}(\bs{\Omega},t)\,\bs{S}_a(0) \label{eq:rotationofsa}
\\
&= \bs{S}_a(0)+\frac{\sin(t|\bs{\Omega}|)}{|\bs{\Omega}|}\,\bs{\Omega}\times \bs{S}_a(0) 
\nonumber
\\
&\quad +\frac12\left(\frac{\sin(\tfrac12 t|\bs{\Omega}|)}{\tfrac12 |\bs{\Omega}|}\right)^2
\bs{\Omega}\times \bs{\Omega}\times \bs{S}_a(0) .
\nonumber
\end{align}
Alternatively, the rotation can be efficiently implemented using quaternions (see, e.g., 
\cite[Sect.\,VII.5.3]{HaiLubWan06}).
No constant rotation vector $\bs{\Omega}$ appears in the spin-orbit and spin-spin Hamiltonians (\ref{eq:hso})-(\ref{eq:hsisi}). However, applying further splittings and appropriate reformulations will nevertheless allow us to make use of the rotation formula for each of them.
We note in passing that although for the full system the spin evolution cannot be determined analytically, a perturbative calculation exists for approximately equal masses for the spin-orbit part~\cite{Tes09}. 

{\bf Spin-orbit integrator.} Up to a constant factor, which we omit in the following, the spin-orbit Hamiltonian (\ref{eq:hso}) is given by
\begin{equation}
 H_\text{SO}(\bs{X},\bs{P},\bs{S}_1,\bs{S}_2) = \bs{S}_{\text{eff}}\cdot \bs{L}/R^3 
\end{equation}
with
$\bs{S}_{\text{eff}}=c_1\bs{S}_1 + c_2\bs{S}_2$.
A way to arrive at a symplectic approximation to its flow, is to 
split 
\begin{equation}
 H_{\text{SO}} = H_{\text{SO}}^1 + H_{\text{SO}}^2 + H_{\text{SO}}^3
\ \text{ with }
 H_{\text{SO}}^i = {S}_{\text{eff}}^i\, L^i/R^3.
\end{equation}
As we show next, the equations of motion for $H_{\text{SO}}^i$ can be solved exactly by rotations of $\bs{X},\bs{P},\bs{S}_1,\bs{S}_2$. Denoting by $\bs{e}_i$ the $i$th unit vector, the equations of motion read (for $i=1$, and analogously for $i=2,3$)
\begin{align}
 \dot{\bs{X}}&= S_{\text{eff}}^1\,\bs{e}_1 \times \bs{X}/R^3
\\
\dot{\bs{P}}&= S_{\text{eff}}^1\,\bs{e}_1\times \bs{P}/R^3 + 3 H_{\text{SO}}^1 \, \bs{X}/R^2
\\
\dot{\bs{S}}_a &= c_a L^1\bs{e}_1\times \bs{S}_a/R^3.
\end{align}
The first equation implies 
$\frac{d}{dt}R^2 =2\dot{\bs{X}}\cdot\bs{X}=0$, so that $R=\,$const. The spin equations imply $\dot S_a^1=0$ and hence $S_{\text{eff}}^1=\,$const. This shows that the equation for $\bs{X}$ is solved by the rotation
${\cal{R}}(S_{\text{eff}}^1\,\bs{e}_1/R^3,t)$.  The equation for $\bs{P}$ is then also solved analytically. For the angular momentum we obtain from the equations of motion for $\bs{X}$ and $\bs{P}$ the differential equation
\begin{equation}
 \dot{\bs{L}}= S_{\text{eff}}^1\,\bs{e}_1\times \bs{L}/R^3,
\end{equation}
which implies ${\dot L}^1=0$, so that also $L^1=\,$const., and hence the spin equations are also solved by simple rotations.
This way, the flow $\varphi_t^{H_{\text{SO}}^1}$ is obtained by four rotations:
\begin{align}
 \bs{X}(t) &= {\cal R}(S_{\text{eff}}^1\,\bs{e}_1/R^3,t)\,
\bs{X}(0)\\
 \bs{P}(t) &= {\cal R}(S_{\text{eff}}^1\,\bs{e}_1/R^3,t)\,
\bigl(\bs{P}(0)
+ 3t H_{\text{SO}}^1 \bs{X}(0)/R^2\bigr)
\\
\bs{S}_a(t) &=  {\cal R}(c_a L^1\,\bs{e}_1/R^3,t)\,
\bs{S}_a(0).
\end{align}
 We then approximate the flow $\varphi_{h/2}^{H_{\text{SO}}}$ by the splitting
\begin{equation}
 \Phi_{h/2}^{H_{\text{SO}}}= \varphi_{h/2}^{H_{\text{SO}}^3}\circ
\varphi_{h/2}^{H_{\text{SO}}^2}\circ \varphi_{h/2}^{H_{\text{SO}}^1}
\end{equation}
or by the adjoint method, with reversed order of the flows,
\begin{equation}
 \Phi_{h/2}^{H_{\text{SO}},\ast}= \varphi_{h/2}^{H_{\text{SO}}^1}\circ
\varphi_{h/2}^{H_{\text{SO}}^2}\circ \varphi_{h/2}^{H_{\text{SO}}^3}.
\end{equation}

{\bf Spin-spin integrator.}
In the following an approximation to the spin-spin coupling flow $\varphi_{h/2}^{H_\text{SS}}$ is derived by a splitting of the spin-spin Hamiltonian, where again each part is readily integrated exactly by rotations. For related splitting approaches for classical spin systems we refer to \cite{FraHuaLei97, McLONe06, SteSch06}, see also \cite{FarLasCou09}.

Each of the terms appearing in the spin-spin Hamiltonian (\ref{eq:hspin}) belongs to one of the forms (omitting constant factors)
\begin{subequations} \label{eq:spinhstructure}
\begin{align}
H^\text{A}&=\bs{S}_1\cdot\bs{S}_2/R^3,\label{eq:spinhstructureb}\\[2pt]
H^\text{B}&=\bs{S}_a\cdot\bs{S}_a/R^3.\label{eq:spinhstructuree}\\[2pt]
H^\text{C}&=(\bs{S}_1\cdot\bs{N})(\bs{S}_2\cdot\bs{N})/R^3,\label{eq:spinhstructurec}\\[2pt]
H^\text{D}&=\tfrac12(\bs{S}_a\cdot\bs{N})^2/R^3,\label{eq:spinhstructured}
\end{align}
\end{subequations}
Since there is no dependence on $\bs{P}$ in any of these Hamiltonians, it follows that
$\bs{X}(t)$ remains constant:
\begin{equation}
 \bs{X}(t) = \bs{X}(0).
\end{equation}
 The dependence on $\bs{X}$ is through the factor $1/R^3$ and in cases C and D also via the unit vector $\bs{N}=\bs{X}/R$. 
In case A we have
${\partial H^{\text{A}}}/{\partial \bs{X}}=-3 H^{\text{A}} \bs{X}/R^2$, 
which is constant along the evolution. 
Therefore, we obtain 
\begin{align}
 \bs{P}(t) &= \bs{P}(0) + 
{3t} \,H^{\text{A}}\,\bs{X}/R^2.
\end{align}
The analogous formula is valid also for $H^{\text{B}}$.
Due to the cross product in (\ref{eq:spineom}), the Hamiltonian $H^{\text{B}}$ does not contribute at all to the motion of the spins. This feature can be exploited for the integration of the spin equations for $H^{\text{A}}$, which read
\begin{align}
\dot{\bs{S}}_1&=\bs{S}_2\times\bs{S}_1/R^3=(\bs{S}_1+\bs{S}_2)\times\bs{S}_1/R^3,
\label{eq:reformulationb1}\\
\dot{\bs{S}}_2&=\bs{S}_1\times\bs{S}_2/R^3=(\bs{S}_1+\bs{S}_2)\times\bs{S}_2/R^3,
\label{eq:reformulationb2}
\end{align}
where consequently the sum $\dot{\bs{S}}_1+\dot{\bs{S}}_2$ is found to be zero. Hence, $\bs{S}_1+\bs{S}_2$ must be constant and Eqs.~(\ref{eq:reformulationb1}) and (\ref{eq:reformulationb2})  are therefore solved by rotation,
\begin{align}
 \bs{S}_a(t)&={\cal R}\bigl((\bs{S}_1(0)+\bs{S}_2(0))/R^3\,,t\bigr)\,\bs{S}_a(0).
\label{eq:rotationofsb}
\end{align}
The Hamiltonian $H^{\text{C}}$ has the spin equations
\begin{align}
 \dot{\bs{S}}_1&=(\bs{S}_2\cdot\bs{N})\,\bs{N}\times\bs{S}_1/R^3,
\\
 \dot{\bs{S}}_2&=(\bs{S}_1\cdot\bs{N})\,\bs{N}\times\bs{S}_2/R^3.
\end{align}
Taking the inner product with $\bs{N}$ shows that both $\bs{S}_1\cdot\bs{N}$ and
$\bs{S}_2\cdot\bs{N}$ are constant. The equations are therefore solved by rotation,
\begin{align}
 \bs{S}_1(t)&={\cal R}\bigl((\bs{S}_2(0)\cdot\bs{N})\bs{N}/R^3\,,t\bigr)\,\bs{S}_1(0)\\
 \bs{S}_2(t)&={\cal R}\bigl((\bs{S}_1(0)\cdot\bs{N})\bs{N}/R^3\,,t\bigr)\,\bs{S}_2(0).
\end{align}
For $H^{\text{C}}$, the equations of motion for the momenta read
\begin{equation}
 \dot{\bs{P}} = 5H^{\text{C}}\,\bs{X}/R^2 - \bs{S}_1(\bs{S}_2\cdot\bs{N})/R^4 -
\bs{S}_2(\bs{S}_1\cdot\bs{N})/R^4,
\end{equation}
which are solved by direct integration:
\begin{align}
 \bs{P}(t) = &\,\bs{P}(0) +5t\,H^{\text{C}}\,\bs{X}/R^2 
\\
&- {\cal Q}\bigl((\bs{S}_2(0)\cdot\bs{N})\bs{N}/R^3,t\bigr)
\bs{S}_1(0)\bigl(\bs{S}_2(0)\cdot\bs{N}\bigr)/R^4
\nonumber
\\
&- {\cal Q}\bigl((\bs{S}_1(0)\cdot\bs{N})\bs{N}/R^3,t\bigr)
\bs{S}_2(0)\bigl(\bs{S}_1(0)\cdot\bs{N}\bigr)/R^4,
\nonumber
\end{align}
where
\begin{align}
&{\cal Q}(\bs{\Omega},t) \bs{S}_a(0) = \int_0^t  {\cal R}(\bs{\Omega},t')\bs{S}_a(0)\,dt'
\\
&\ \ \ = t\bs{S}_a(0) + \frac12\left(\frac{\sin(\tfrac12 t|\bs{\Omega}|)}{\tfrac12 |\bs{\Omega}|}\right)^2
\bs{\Omega}\times  \bs{S}_a(0) 
\nonumber
\\
&\ \ \ \ \ \ + \frac{t|\bs{\Omega}|-\sin(t|\bs{\Omega}|)}{|\bs{\Omega}|^3}\,
\bs{\Omega}\times \bs{\Omega}\times \bs{S}_a(0).
\nonumber
\end{align}
The treatment of the Hamiltonian $H^\text{D}$ is similar to the just
expounded procedure and is therefore not shown. Taking all parts together,  we approximate the spin-spin flow $\varphi_{h/2}^{H_{\text{SS}}}$  by the composition of the exact flows of the subsystems, written schematically as
\begin{align}
 \Phi_{h/2}^{H_{\text{SS}}} &=\varphi_{h/2}^\text{D}\circ
\varphi_{h/2}^\text{C}\circ \varphi_{h/2}^\text{B} \circ \varphi_{h/2}^\text{A}.
\end{align}
Its adjoint $\Phi_{h/2}^{H_{\text{SS}},\ast}$ features the reversed order of all contributions. We have chosen an ordering of the various sub-Hamiltonians which might not be optimal. We note, however, that the flows of A and B commute, since $\{ H^\text{A}, H^\text{B}\}=0$. This implies that no splitting error occurs between A and B. Similarly, the flows of C and D also commute. As compositions of exact flows, the discrete spin-spin flows 
$\Phi_{h/2}^{H_{\text{SS}}}$ and $\Phi_{h/2}^{H_{\text{SS}},\ast}$ are Poisson integrators.

{\bf Higher order by composition.} The overall method 
\begin{equation}
\varphi_h^H\approx\Phi_h^H=
\Phi_{h/2}^{H_{\text{SS}},\ast}\circ 
\Phi_{h/2}^{H_{\text{SO}},\ast}\circ 
\Phi_{h}^{H_{\text{Orb}}}\circ 
\Phi_{h/2}^{H_{\text{SO}}}\circ 
\Phi_{h/2}^{H_{\text{SS}}}
\label{eq:splittingallapprox}
\end{equation}
is a Poisson integrator as a composition of Poisson maps. It is only of second order, but its order can be enhanced by composition; see, e.g., \cite[Chap.\,V.3]{HaiLubWan06} or \cite{McLQui02}. In our numerical experiments we will lift its order to four by Suzuki's composition scheme \cite{Suz90}:
\begin{align}
\Phi_h^{H,\,\text{4th order}}=\Phi_{\gamma_5 h}^H
\,\circ\,\Phi_{\gamma_4 h}^H \,\circ\,\Phi_{\gamma_3 h}^H \,\circ\,\Phi_{\gamma_2 h}^H \,\circ\,\Phi_{\gamma_1 h}^H,
\label{eq:suzukiflow}
\end{align}
where the lengths of the substeps are
\begin{equation}
\gamma_1=\gamma_2=\gamma_4=\gamma_5=\frac{1}{4-4^{1/3}}\,, ~~~\gamma_3=-\frac{4^{1/3}}{4-4^{1/3}}.
\end{equation}
The just introduced integration scheme Eq.~(\ref{eq:splittingallapprox}) shall be termed SPN integrator (symplectic post-Newtonian integrator). The fourth-order version Eq.~(\ref{eq:suzukiflow}) will be referred to as SPN4 in the following.


\section{Numerical Experiments}
\label{sec:numresults}

For the demonstration of the properties and performance of the symplectic integrator we use the following configuration, consisting of two black holes with non-trivial, maximal spin vectors:
\begin{align}
m_1 &= 0.25\nonumber\\
m_2 &= 0.75\nonumber\\
R   &= (50, 0, 0)\nonumber\\
P   &= (0, 0.027475637, 0)\label{eq:initdata}\\
S_1  &= m_1^2\cdot(-1, 0, 0)\nonumber\\
S_2  &= m_2^2\cdot(  1/\sqrt{2},~~0,~~1/\sqrt{2})\nonumber
\end{align}
These initial data yield low-eccentricity orbits (see \cite{WalBruMue09}), the average period of which is approximately $T\approx 2,286$.

For comparison with the newly developed fourth-order symplectic integrator we use the classical fourth-order Runge-Kutta method (see, e.g., \cite{HaiNorWan93}) in a self-made C version. Inadequacies in the efficiency of the implementation of either the RK method or the symplectic scheme would only affect the CPU timings of the subsequent results but not the principal behavior and insights.

\begin{figure}[ht!]
\begin{overpic}[width=8.3cm]{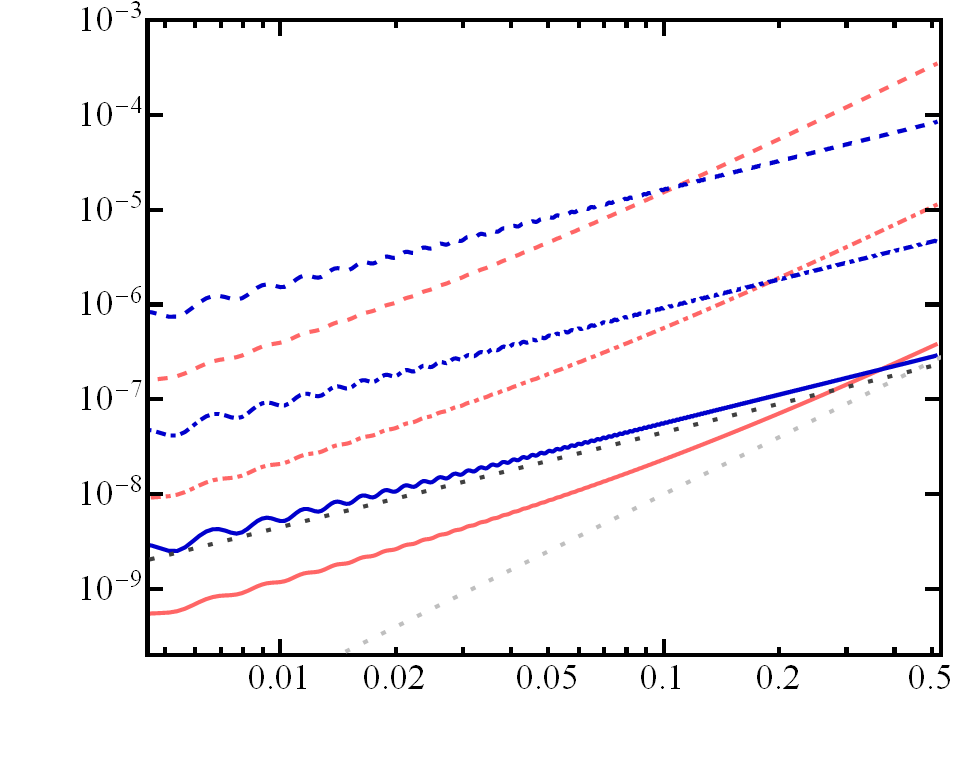}
\put(2,44){\begin{sideways}$\varepsilon$\end{sideways}}
\put(50,3){$t~/~10^{6}$}
\end{overpic}
\begin{tabular}{lcccccclccccc}
~& {\scriptsize Method} & &{\scriptsize $h$} & ~~& {\scriptsize $t_\text{CPU}$} & ~~~~~~ & ~~& {\scriptsize Method} &~~ &  {\scriptsize $h$}&~~ & {\scriptsize $t_\text{CPU}$}\\%
\includegraphics[width=0.6cm]{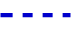} &\hskip0.01cm {\scriptsize SPN4} & & {\scriptsize $64$} & &  {\scriptsize $\phantom{1}5.1\,\text{s}$}%
& ~~&\includegraphics[width=0.6cm]{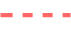} &\hskip0.01cm {\scriptsize RK4}& & {\scriptsize $8$} & &  {\scriptsize $\phantom{1}9.1\,\text{s}$} \\%
\includegraphics[width=0.6cm]{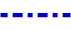} &\hskip0.01cm {\scriptsize SPN4} & & {\scriptsize $32$}& &   {\scriptsize $10.2\,\text{s}$}%
& &\includegraphics[width=0.6cm]{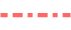} &\hskip0.01cm {\scriptsize RK4}&  & {\scriptsize $4$}& &  {\scriptsize $18.5\,\text{s}$} \\%
\includegraphics[width=0.6cm]{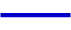} &\hskip0.01cm {\scriptsize SPN4} & & {\scriptsize $16$} & &  {\scriptsize $20.8\,\text{s}$}%
& &\includegraphics[width=0.6cm]{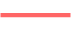} &\hskip0.01cm {\scriptsize RK4} & & {\scriptsize $2$} & & {\scriptsize $37.1\,\text{s}$} \\%
\includegraphics[width=0.6cm]{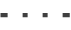} &\multicolumn{5}{l}{\hskip0.01cm \scriptsize{~~Linear growth}}%
& &\includegraphics[width=0.6cm]{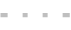} &\multicolumn{5}{l}{\hskip0.01cm \scriptsize{~~Quadratic growth}} \\
\end{tabular}
\caption{Comparison of error growth between SPN4 and RK4 method for different step sizes.}
\label{fig:conservativeerrorgrowth}
\end{figure}

We measure the accuracy of the numerical integration by the scaled error norm
\begin{align}
\varepsilon=\left[\left(\frac{\Delta X}{X}\right)^2+\left(\frac{\Delta P}{P}\right)^2+\left(\frac{\Delta S_1}{S_1}\right)^2+\left(\frac{\Delta S_2}{S_2}\right)^2\right]^{1/2},
\end{align}
where the individual terms are computed as the Euclidean norm of the componentwise deviation from a reference solution (RK4, step size $h=0.0125$), for instance,
\begin{align}
\frac{\Delta X}{X}=\left[\sum_{i=1}^3 \left(\frac{X_i-X_i^\text{ref}}{X_i^\text{ref}} \right)^2 \right]^{1/2}.
\end{align}

In the SPN4 method there is a fixed-point iteration to solve the implicit equations (\ref{eq:phipn})-(\ref{eq:phipnadjoint}). As a stopping criterion for this iteration we use the condition
\begin{align}
\left[\sum_{i=1}^3 \left(X_i^n-X_i^{n-1}\right)^2~ /~ \sum_{i=1}^3 \left(X_i^{n}\right)^2 \right]^{1/2}~&<~\epsilon_\text{FPI}
\end{align}
for Eq.~(\ref{eq:phipn}), and
\begin{align}
\left[\sum_{i=1}^3 \left(P_i^n-P_i^{n-1}\right)^2~ /~ \sum_{i=1}^3 \left(P_i^{n}\right)^2 \right]^{1/2}~&<~\epsilon_\text{FPI}
\end{align}
for Eq.~(\ref{eq:phipnadjoint}). The number of iterations will be limited to $N_\text{FPI}$. In the comparisons below we chose 
the combination $\epsilon_\text{FPI}=10^{-12}$ and $N_\text{FPI}=5$, which showed appropriate performance for our precision demands.
The step size $h$ is chosen to give roughly comparable computer time
for the two methods, implying a larger $h$ for SPN4 in our examples.

As indicated in Fig.~\ref{fig:conservativeerrorgrowth} the error growth of the symmetric, symplectic method SPN4 is linear, a feature which is not unexpected (cf.~\cite[Chaps.\, X and XI]{HaiLubWan06}). By contrast, the non-symmetric, non-symplectic RK method suffers from quadratic error growth, making it inferior for long-term simulations. In the example the simulation time is $t=512,000$ which gives about 224 revolutions of the binary black-hole system. Supposing approximately equal numerical costs, there evidently exists a time $t_\text{break even}$ beyond which the SPN4 method gives more accurate solutions than RK4. An analysis of the CPU time spent for the different Hamiltonians in the SPN4 method reveals that 
about 88\% is consumed by the $H_\text{PN}$ part, 9\% by $H_\text{N}$, and only 3\% for all the spin parts together.

Fig.~\ref{fig:conservativeerrorgrowth} also depicts the 4th-order convergence of both methods, SPN4 and RK4. Dividing the step size by two yields an error which is smaller by a factor 1/16.

\begin{figure}[ht!]
\begin{overpic}[width=8.3cm]{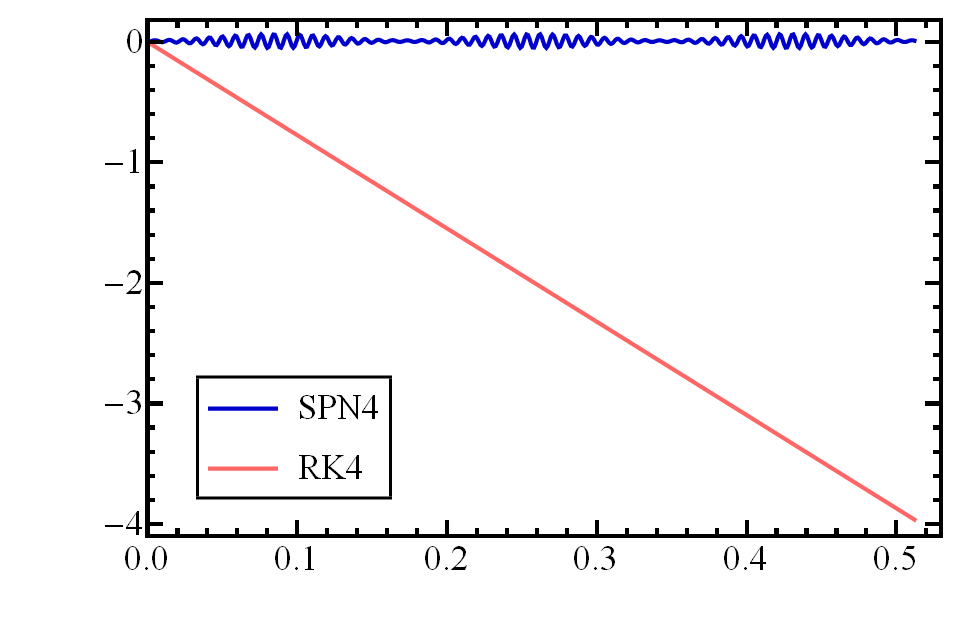}
\put(2,25){\begin{sideways}$E-E_0~/~10^{-10}$\end{sideways}}
\put(50,3){$t~/~10^{6}$}
\end{overpic}
\caption{The symplectic method ($h=64$) preserves a nearby Hamiltonian, whereas the RK integrator ($h=8$) produces a drift. The system's exact energy is about $E_0\approx-1.817\cdot 10^{-3}$.}
\label{fig:conservativeenergy}
\end{figure}

\begin{figure}[ht!]
\begin{overpic}[width=8.3cm]{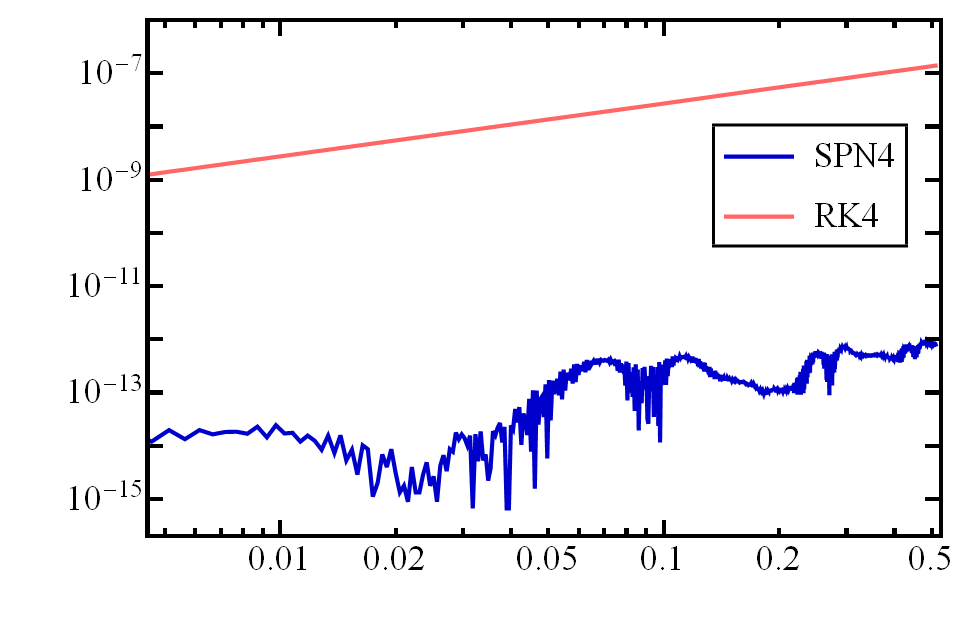}
\put(2,30){\begin{sideways}$|J-J_0|$\end{sideways}}
\put(50,3){$t~/~10^{6}$}
\end{overpic}
\caption{The variations of $J$ induced by the symplectic method can be attributed to machine-precision effects.}
\label{fig:conservativej}
\end{figure}

\begin{figure}[ht!]
\begin{overpic}[width=8.3cm]{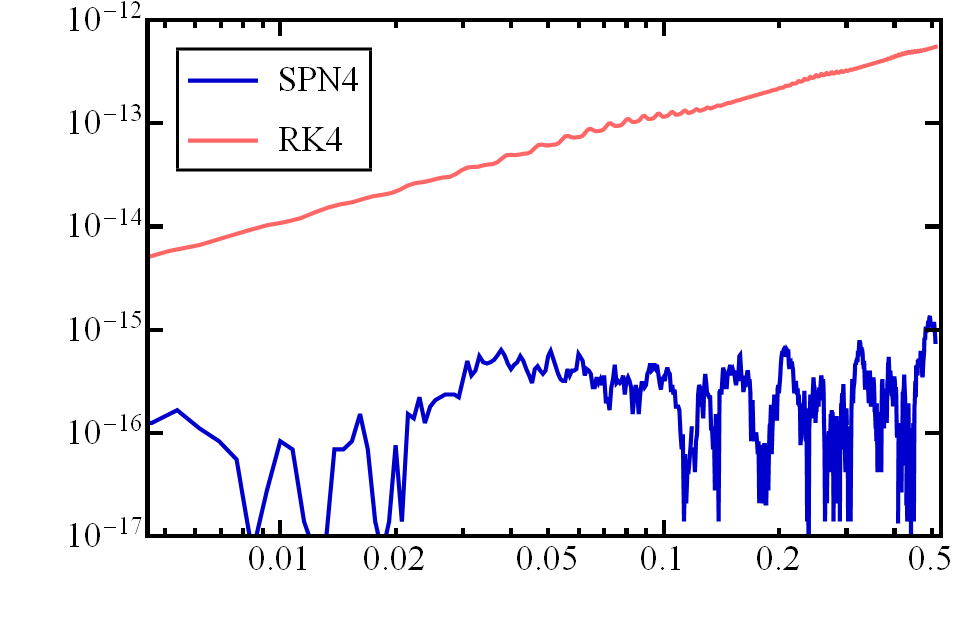}
\put(2,30){\begin{sideways}$|S_1-S_1^0|$\end{sideways}}
\put(50,3){$t~/~10^{6}$}
\end{overpic}
\caption{The spin lengths are preserved exactly by the SPN method. Oscillations are due to machine-precision effects. The RK integrator yields a drift.}
\label{fig:conservatives1}
\end{figure}

Another striking feature of the symplectic integrator is the near-conservation of the energy of the system. Fig.~\ref{fig:conservativeenergy} shows that only minor oscillations in the system's energy occur, whereas a linear drift results for the non-symplectic RK method, becoming severe for long-term simulations. A similar observation is made for the other conserved quantities $J$ and $S_a$, that is the absolutes of the total angular momentum and the spins, cf. Figs.~\ref{fig:conservativej} and \ref{fig:conservatives1}.

\section{Adding Dissipation}
\label{sec:dissipation}

So far the non-conservative term $\bs{F}$ in Eq.~(\ref{eq:nonconservp}) has been omitted, and the construction of the SPN method was targeted to the conservative equations. However, in the physically realistic scenario dissipation has to be accounted for. Energy is transferred into propagating curvature of spacetime, the radiation of gravitational waves.

As is known from \cite[Chap.\,XII]{HaiLubWan06}, it is advantageous for dissipatively perturbed systems to use numerical methods that reduce to a symmetric and/or symplectic method for the unperturbed, conservative system. Therefore we use the SPN method as derived so far and incorporate the dissipative force $\bs{F}$ via a further splitting. The starting point is the symplectic flow before it is lifted from second to fourth order by the Suzuki composition scheme (\ref{eq:suzukiflow}). The dissipative contribution, denoted by $\varphi_{h/2}^F$, is padded around the conservative flow,
\begin{align}
\varphi_h^\text{diss}\approx\varphi^{F}_{h/2}\,\circ\,\varphi_{h}^{\text{cons}}\,\circ\,\varphi^F_{h/2}.
\label{eq:dissipativesplitting}
\end{align}
In the numerical implementation the conservative flow is again approximated by Eq.~(\ref{eq:splittingallapprox}), and the dissipative perturbation is modeled by an Euler step
\begin{align}
 \Phi_h^F:\ 
\widehat{\bs{P}} &= \bs{P} + \frac h2 \bs{F}(\bs{X},\bs{P},\dot{\bs{X}})\ \text{ with }\ 
\dot{\bs{X}}=\parder{H}{\bs{P}}(\bs{X},\bs{P}).
\end{align}
The adjoint method is the implicit Euler method,
\begin{align}
 \Phi_h^{\ast\, F}:\ 
\widetilde{\bs{P}} &= \bs{P} + \frac h2 \bs{F}(\bs{X},\widetilde{\bs{P}},\widetilde{\dot{\bs{X}}})\ \text{ with }\ 
\widetilde{\dot{\bs{X}}}=\parder{H}{\bs{P}}(\bs{X},\widetilde{\bs{P}}).
\end{align}
This results in a symmetric, second-order integrator given by
\vspace{-0.3cm}
\begin{align}
\Phi_h^\text{diss} &= \Phi^{\ast \,F}_{h/2}\,\circ\,\Phi_{h}^{\text{cons}}\,\circ\,\Phi^F_{h/2}.\label{eq:dissipativesplittingapprox}
\end{align}
Note that we have to provide the computationally expensive term $\partial{H}/\partial{\bs{P}}$ in order to compute the force $\bs{F}$. Retaining the symmetry of Eq.~(\ref{eq:dissipativesplittingapprox}) is, however, of no structural importance for the dissipative contributions. Hence, if we replace the implicit Euler method by the explicit method resulting from two fixed-point iterations, we obtain overall a second-order method that uses three additional evaluations of $\partial{H}/\partial{\bs{P}}$ to incorporate the dissipative forces. Applying Suzuki composition, Eq.~(\ref{eq:suzukiflow}), the order of the method then is increased to four (or higher).

\begin{figure}[ht!]
\vskip-0.2cm
\begin{overpic}[width=8.3cm]{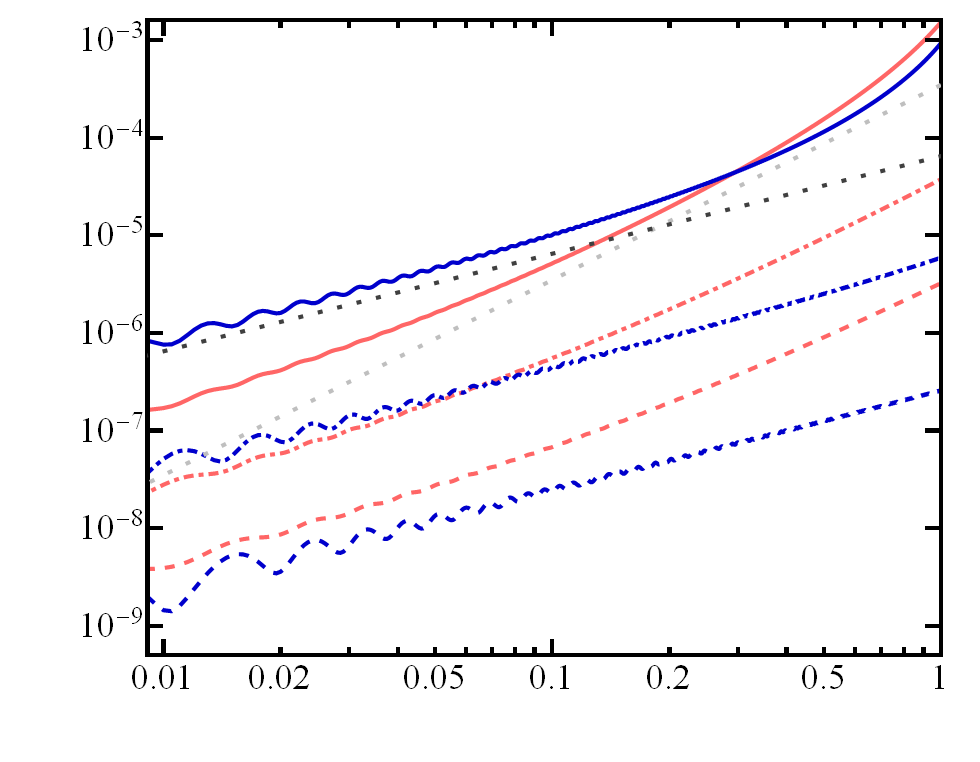}
\put(2,44){\begin{sideways}$\varepsilon$\end{sideways}}
\put(50,3){$t/t_\text{max}$}
\end{overpic}
\begin{overpic}[width=8.3cm]{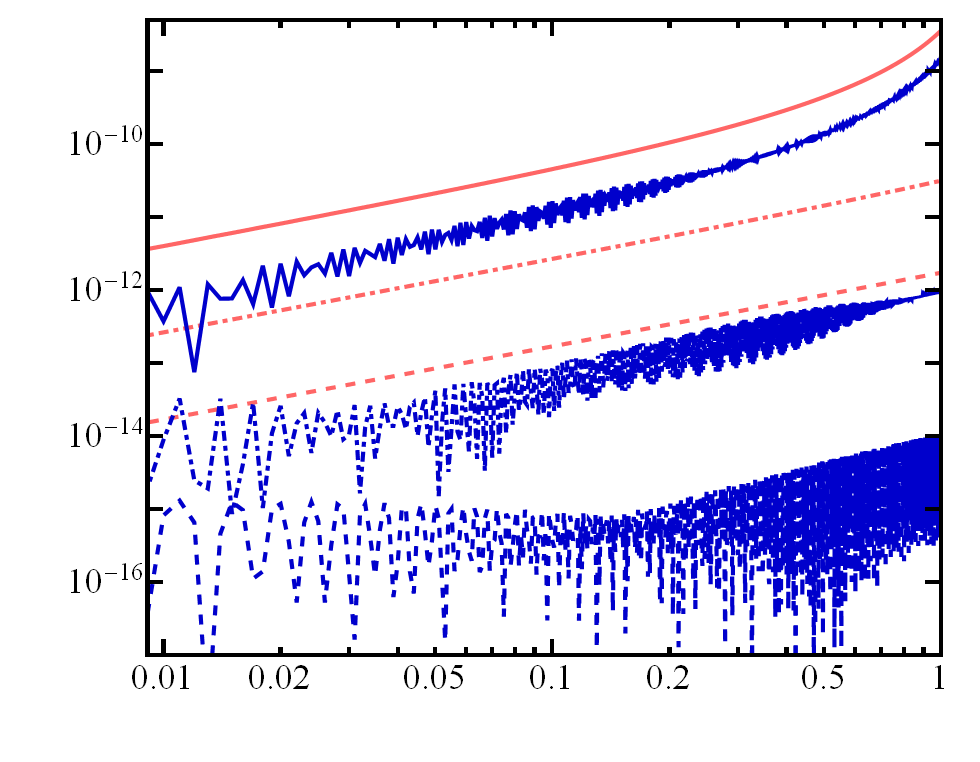}
\put(2,34){\begin{sideways}$|E-E^\text{ref}|$\end{sideways}}
\put(50,3){$t/t_\text{max}$}
\end{overpic}
\begin{tabular}{lccccclcccc}
~& {\scriptsize Method} & {\scriptsize $R_\text{start}$} & {\scriptsize $h$} & {\scriptsize $t_\text{CPU}$} & ~~~~ & & {\scriptsize Method} & {\scriptsize $R_\text{start}$} & {\scriptsize $h$} & {\scriptsize $t_\text{CPU}$}\\%
\includegraphics[width=0.6cm]{figures/spn_dashed.png} &\hskip0.01cm {\scriptsize SPN4}& {\scriptsize $200$}  & {\scriptsize $256$} &  {\scriptsize $\phantom{1}6.1\,\text{s}$}%
& &\includegraphics[width=0.6cm]{figures/rk_dashed.png} &\hskip0.01cm {\scriptsize RK4} & {\scriptsize $200$} & {\scriptsize $32$} &  {\scriptsize $10.3\,\text{s}$} \\%
\includegraphics[width=0.6cm]{figures/spn_dotdashed.png} &\hskip0.01cm {\scriptsize SPN4}& {\scriptsize $100$}  & {\scriptsize $128$} &  {\scriptsize $\phantom{1}6.3\,\text{s}$}%
& &\includegraphics[width=0.6cm]{figures/rk_dotdashed.png} &\hskip0.01cm {\scriptsize RK4} & {\scriptsize $100$} & {\scriptsize $16$} &  {\scriptsize $10.1\,\text{s}$} \\%
\includegraphics[width=0.6cm]{figures/spn_solid.png} &\hskip0.01cm {\scriptsize SPN4}& {\scriptsize $\phantom{1}50$}  & {\scriptsize $\phantom{1}64$} &  {\scriptsize $\phantom{1}6.6\,\text{s}$}%
& &\includegraphics[width=0.6cm]{figures/rk_solid.png} &\hskip0.01cm {\scriptsize RK4} & {\scriptsize $\phantom{1}50$} & {\scriptsize $\phantom{1}8$} &  {\scriptsize $10.3\,\text{s}$} \\%
\includegraphics[width=0.6cm]{figures/darkgray_sparsedotted.png} &\multicolumn{4}{l}{\hskip0.01cm \scriptsize{~~Linear growth}}%
& &\includegraphics[width=0.6cm]{figures/gray_sparsedotted.png} &\multicolumn{4}{l}{\hskip0.01cm \scriptsize{~~Quadratic growth}} \\
\end{tabular}
\vskip-0.1cm
\caption{Error growth and energy deviation for the dissipative implementation of SPN4 and the RK4 method starting from different initial separations $R_\text{start}$. 
Applying different step sizes while keeping the number of steps (8000 for SPN4, 64000 for RK4) results in different simulation times $t_\text{max}$, used for normalization of the time axis.}
\label{fig:dissipativeerrorgrowthvariousradii}
\end{figure}

For the numerical experiments the initial data Eq.~(\ref{eq:initdata}) are modified to still give low-eccentricity orbits in the dissipative regime. For example, an inspiral starting at $R=50M$ would have the initial momentum
\begin{align}
P   &= (-3.5267394\cdot 10^{-6}, 0.027475637, 0),\label{eq:initdatadiss}
\end{align}
obtained with the algorithm presented in \cite{WalBruMue09}. The energy loss in the system now causes the separation of the black holes to decline from $R=50$ down to $R=36.7$ during the simulation. In general, there would be a need for stepsize control mechanisms, especially for longterm simulations. A reversible step size strategy as in \cite[Sect.\,VIII.3.2]{HaiLubWan06} can be employed based on the conservative part, but this is beyond the scope of the present work. In our numerical experiments, a constant stepsize is chosen such that sufficient resolution is granted throughout the simulation. 

Variation of the initial separation $R_\text{start}$ provides a
possibility to adjust the degree of dissipation. For a binary
separation of $50M$ the perturbation is already quite strong. The
ratio of dissipative force to conservative terms amounts to
$|\bs{F}|/|\partial{H}/\partial{\bs{X}}|\approx 6.0\cdot10^{-5}$.
For separations of $100M$ and $200M$ the ratio goes down to
$1.1\cdot10^{-5}$ and $2.0\cdot10^{-6}$,
respectively. Fig.~\ref{fig:dissipativeerrorgrowthvariousradii} shows
how increasing dissipation affects the performance of the
integrators. The linear error growth of the SPN4 integrator does not
persist for strong dissipation. However, its performance remains
superior to the RK4 method, which also detaches from the quadratic
error growth. In a similar fashion the ideal behavior of the SPN4
method concerning energy and angular momentum, i.e.\ minor
fluctuations around the exact values, is lost. Even for very small
perturbation with $R_\text{start}=200M$ we observe a linear growth
exceeding the machine noise. For even stronger dissipation the growth
turns out to be faster than linear. The very same behavior is seen for
the angular momentum (not shown). The spin lengths still remain
preserved by the SPN4 method, due to the construction of the algorithm.
In our numerical experiments the SPN4 method still shows a better performance than the RK4 method for sufficiently long integration intervals.


\section{Conclusion}
\label{sec:conclusion}

We presented a new numerical scheme designed for integrating the
post-Newtonian equations of motion for a spinning black-hole binary.
The special structure of the Hamiltonian enabled us to use a splitting
algorithm that exactly preserves the non-canonical symplectic
structure of the system. The result is an integrator with excellent
long-term performance. The integrator was also modified to include
dissipative effects contained in the more realistic 3.5PN accurate
model. Even though symplecticity is no longer present in the dissipative case, the
integrator was shown to exhibit favorable properties in long-term
simulations with small dissipative forces. With these properties we
expect this new integrator to be a useful and efficient tool in the
investigation of chaotic traits in the dynamics of spinning binary
systems, significantly reducing the risk of numerical artifacts compared to
standard integrators.


\acknowledgments
It is a pleasure to thank Achamveedu Gopakumar and Gerhard Sch\"{a}fer for discussions and valuable insights into
the PN method.
This work was supported in part by DFG grant SFB/Transregio~7
``Gravitational Wave Astronomy'' and the DLR (Deutsches Zentrum f\"ur Luft
und Raumfahrt).
%


\bibliography{spinrefs}

\end{document}